\documentclass[fleqn,usenatbib]{mnras}

\DeclareRobustCommand{\VAN}[3]{#2}
\let\VANthebibliography\thebibliography
\def\thebibliography{\DeclareRobustCommand{\VAN}[3]{##3}\VANthebibliography}

\usepackage[T1]{fontenc}
\usepackage{amsfonts}
\usepackage{amsmath}	
\usepackage{booktabs}
\usepackage{cancel}
\usepackage{color}
\usepackage{epsfig}
\usepackage{epstopdf}
\usepackage{float}
\usepackage{graphicx}
\usepackage{hyperref}
\usepackage{mathrsfs}
\usepackage{multirow}
\usepackage{natbib}
\usepackage{newtxtext,newtxmath}
\usepackage{orcidlink}
\usepackage{subfigure}
\usepackage{times}
\usepackage{ulem}
\usepackage{xcolor}

\usepackage{lineno}
\usepackage{etoolbox}

\newcommand{\patchAmsMathEnvironment}[1]{%
	\expandafter\let\csname old#1\expandafter\endcsname\csname #1\endcsname
	\expandafter\let\csname oldend#1\expandafter\endcsname\csname end#1\endcsname
	\renewenvironment{#1}%
	{\linenomath\csname old#1\endcsname}%
	{\csname oldend#1\endcsname\endlinenomath}%
}
\newcommand{\patchBothAmsMathEnvironments}[1]{%
	\patchAmsMathEnvironment{#1}%
	\patchAmsMathEnvironment{#1*}%
}
\AtBeginDocument{%
	\patchBothAmsMathEnvironments{equation}%
	\patchBothAmsMathEnvironments{align}%
	\patchBothAmsMathEnvironments{flalign}%
	\patchBothAmsMathEnvironments{alignat}%
	\patchBothAmsMathEnvironments{gather}%
	\patchBothAmsMathEnvironments{multline}%
}

\title[Depolarization induced by rapid PA swings]{Depolarization Induced by Rapid Polarization Angle Swings: A Common Feature of Pulsars and Fast Radio Bursts?}

\author[Y.-C. Huang et al.]{
Yu-Chen Huang \orcidlink{0009-0009-7749-8998},$^{1,2}$\thanks{E-mail: ychuang@mail.ustc.edu.cn}
Jie-Shuang Wang \orcidlink{0000-0002-2662-6912},$^{3}$
Song-Bo Zhang \orcidlink{0000-0003-2366-219X} $^{4}$
and Zi-Gao Dai \orcidlink{0000-0002-7835-8585} $^{1,2}$\thanks{E-mail: daizg@ustc.edu.cn}
\\
$^{1}$Department of Astronomy, School of Physical Sciences, University of Science and Technology of China, Hefei 230026, China\\
$^{2}$School of Astronomy and Space Science, University of Science and Technology of China, Hefei 230026, China\\
$^{3}$Tsung-Dao Lee Institute \& School of Physics and Astronomy, Shanghai Jiao Tong University, Shanghai 201210, China\\
$^{4}$Purple Mountain Observatory, Chinese Academy of Sciences, Nanjing 210023, China
}

\date{Accepted XXX. Received YYY; in original form ZZZ}

\pubyear{2026}

\begin{document}
\label{firstpage}
\pagerange{\pageref{firstpage}--\pageref{lastpage}}
\maketitle

\begin{abstract}
The polarization angle (PA) of pulsars and fast radio bursts (FRBs) provides a useful diagnostic of the magnetic fields in their emission regions and is therefore crucial for understanding their radiation and origins. Within a general geometric framework for polarized emission from a rotating neutron star, we suggest a possible anti-correlation between the degree of linear polarization $\Pi_\text{L}$ and $d\text{PA}/dt$, the time derivative of the PA, as a common feature of pulsars and FRBs. The depolarization arises from the incoherent superposition of radiation with different polarization directions within the observable part of the emission region, and is detectable only when the PA swing is steep enough. We test this conjecture using a sample of radio pulsars and find possible evidence for the expected anti-correlation in a subset of pulsars. Whether this relation holds in FRBs remains uncertain due to limited observational data. Identification of this feature would not only provide insights into the rotating magnetospheric origin of FRBs, but also place constraints on the spin periods and geometric parameters of the neutron stars that power these mysterious bursts.
\end{abstract}


\section{Introduction}

Fast radio bursts (FRBs) are mysterious, transient radio phenomena, characterized by extremely short durations and exceptionally high energy releases, with their origins still unknown \citep{Cordes2019,Petroff2019,Xiao2021,Zhang2023}. One widely discussed theoretical postulate is that FRBs originate from the rotating magnetospheres of neutron stars \citep[e.g., ][]{Huang2024,Huang2024a,Liu2025a,Voisin2025,Nishiura2026,Qu2026,Solanki2026}. The polarization angles (PAs) of some FRBs exhibit swings similar to those seen in radio pulsars, which is often regarded as evidence for a rotating magnetospheric origin \citep{Luo2020,Jiang2022,Zhang2023a,Pandhi2024,Liu2025,Mckinven2025,Xie2025,Manaswini2026,Uttarkar2026,Wang2026}. To explain these PA swings, the rotating vector model (RVM), commonly used in pulsars, is often invoked in FRBs \citep{Pandhi2024,Mckinven2025,Liu2025}. The underlying principle of the RVM is that the PA traces the geometric projection of the local magnetic field line in the emission region onto the sky plane, and its variation is caused by the line of sight (LOS) sweeping across the rotating magnetosphere \citep{Radhakrishnan1969}. 
Based on a dipole magnetic field configuration, the RVM successfully explains the characteristic S-shaped PA curves observed in a fraction ($20\mathord{-}30\%$) of pulsars \citep{Everett2001,Johnston2023,Wang2023}.

However, when applied to FRBs, the RVM encounters several challenges. For example, the PA swings in FRBs are often irregular and diverse, and exhibit significant variability from burst to burst \citep{Cho2020,Luo2020,Jiang2022,Zhang2023a,Bera2024,Pandhi2024,Bera2025,Xie2025,Manaswini2026}. Moreover, even for different bursts from the same source, the geometric parameters inferred from tentative RVM fits are not mutually consistent \citep{Liu2025}. One plausible interpretation is that the PA swings reflect highly irregular magnetic field geometries within the emission region that vary from burst to burst. These magnetic field structures deviate significantly from an ideal dipole configuration. Searching for periodicity associated with neutron star spin in FRBs has hitherto yielded no compelling results \citep{Zhang2018,Niu2022,Du2024,Du2025,Gazith2025,Katz2025,Zhou2025}. This seems to further support the idea that the origin of FRBs requires additional explanations beyond a simple dipole configuration. However, other possibilities responsible for the PA swings cannot be completely ruled out. For example, propagation effects in the magnetosphere or near-source plasma may also modulate the PAs of FRBs \citep{Lu2019,Liu2024,Bera2025,Li2026}.

\begin{figure*}
	\centering
	\includegraphics[width=0.9\textwidth]{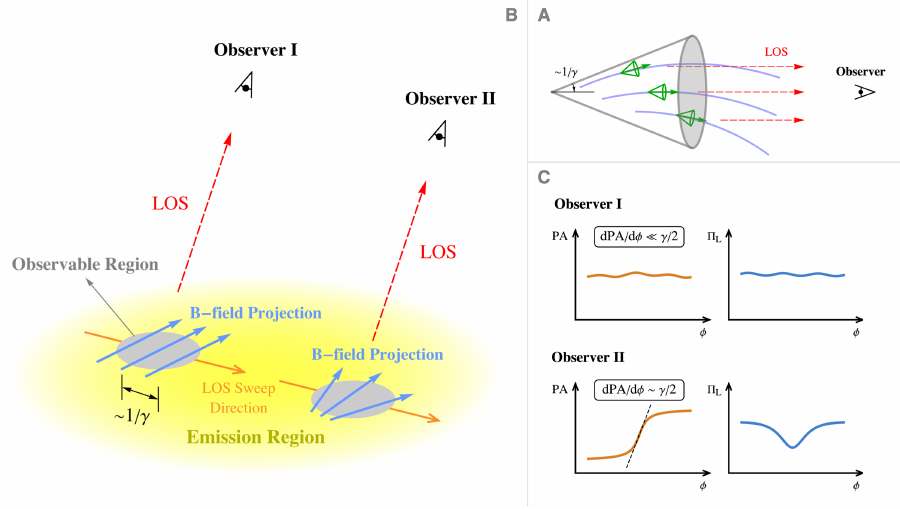}
	\caption{Schematic illustration showing that geometric depolarization could occur when the PA varies rapidly. \textit{Panel A}: The blue curves represent the trajectories of relativistic electrons. Radiation from an electron can be observed only when its velocity direction (green arrow) lies within a cone (gray cone) with a half-opening angle of $\sim1/\gamma$ centered on the LOS (red dashed arrows). Thus, the observed radiation consists of contributions from electrons moving along different adjacent magnetic field lines. \textit{Panel B}: The blue arrows denote the projections of the magnetic field lines in the emission region (yellow region) onto the plane of the sky. The orange arrows indicate the sweep direction of the LOS. The gray disks represent the observable regions for two different observers. These observable regions are deliberately exaggerated to illustrate their internal magnetic field lines. Depolarization occurs when the observable region contains magnetic field lines with different orientations, i.e., when the observed radiation consists of components with different polarization directions. \textit{Panel C}: The predicted PA and degree of linear polarization $\Pi_\text{L}$ as functions of pulse longitude for the two observers. Observer I sees a slow change in PA and little or no geometric depolarization, whereas Observer II sees a rapid change in PA and significant geometric depolarization.}
	\label{depoillus}
\end{figure*}

In this paper, we investigate whether the PA swings in FRBs are caused by a geometric effect similar to that in the RVM. We note that a natural outcome of this geometric effect is that depolarization could occur when the magnetic field orientations in the emission region are disordered. Previous studies have suggested that this phenomenon may occur when the line of sight (LOS) approaches the magnetic pole of a pulsar, due to the diverging magnetic field lines in the polar region \citep{Gil1985,Gil1985a}. In fact, a more general conjecture is that sufficiently rapid PA swings may lead to significant depolarization. We provide a schematic illustration of this geometric depolarization phenomenon in Fig. \ref{depoillus}, which is important for the subsequent discussion. Due to the relativistic beaming effect, the radiation from a particle predominantly confined within a cone centered along its direction of velocity, with a half-opening angle $\sim1/\gamma\ll1$, where $\gamma$ is the Lorentz factor of the particle \citep{Rybicki1991}. The emission is observable only when the LOS falls within the cone, or equivalently, when the velocity vector of a particle lies within a cone of the same opening angle centered on the LOS. Such an extremely narrow cone determines the observable part of the whole emission region. Within the small observable region, the magnetic field lines projected onto the plane of the sky are well ordered in most cases, as reflected by the observed highly polarized radio radiation from neutron stars. However, the situation changes when the PA varies extremely rapidly: the orientations of the magnetic field lines within the observable region are no longer aligned, which then produces depolarized radiation.

We note that such a depolarization phenomenon primarily originates from the geometric effect rather than the specific radiation mechanism. Therefore, we speculate that a similar process may operate in FRBs. Moreover, an advantage of this geometric depolarization mechanism is that it does not necessarily require the magnetic field to be dipolar. As long as the PA varies sufficiently rapidly, significant depolarization can naturally occur. Such a feature is expected when the emission region exhibits transverse motion relative to the LOS. It could therefore serve as a probe of the rotating magnetospheric origin of FRBs. We further predict that depolarization would occur when the PA varies so rapidly that it exhibits a significant change within the observable region:
\begin{equation}
	\Delta\text{PA}\sim\frac{d\text{PA}}{d\phi}\Delta\phi\sim\frac{d\text{PA}}{d\phi}\frac{2}{\gamma}\sim1\Longrightarrow\frac{d\text{PA}}{d\phi}\sim\frac{\gamma}{2},
	\label{criterion}
\end{equation}
where $d\text{PA}/d\phi$ (hereafter, we ignore its sign for convenience) is defined as the derivative of the PA with respect to the spin phase $\phi$. $\Delta\phi\sim2/\gamma$ is the phase width occupied by the observable region, as shown in Fig. \ref{depoillus}. Equation (\ref{criterion}) provides a rough criterion for the depolarization induced by rapid PA swings. In Section \ref{theory}, we present a more detailed discussion and quantitatively describe the relation between the degree of linear polarization and the rate of change of the PA, based on a simplified model. We also test this relation in a sample of pulsars in Section \ref{testpulsar} to assess its plausibility. In Section \ref{imfrbs}, we demonstrate how this relation can be used to constrain the spin periods and geometric parameters of FRB progenitors, assuming that FRBs originate from rotating neutron stars. This could also be used to examine recent models which suggest repeating FRBs originate from magnetars with aligned spin and magnetic axes \citep{Beniamini2025,Luo2025,Rajwade2025,Zhang2025}. The implications are discussed in Section \ref{discussion}.

\section{Model}\label{theory}

Given our limited understanding of the radiation process of FRBs, we attempt to proceed without specifying a particular emission mechanism. We consider a relativistic emitter (the smallest coherent unit), denoted by $i$, with a Lorentz factor $\gamma$, that emits a radiation beam. Its Stokes parameters can be expressed by \citep{Rybicki1991}
\begin{equation}
	\begin{aligned}
		I_i&=E_i^2,\quad Q_i=E_i^2 \cos{2\chi_i}\cos{2\psi_i},\\
		U_i&=E_i^2 \cos{2\chi_i}\sin{2\psi_i},\quad  V_i=E_i^2 \sin{2\chi_i},
	\end{aligned}
\end{equation}
where $E_i$ is the electric field strength, $\chi_i$ and $\psi_i$ are the ellipticity angle and the PA, respectively. A key ingredient of our model is that the observed radiation consists of contributions from a large number ($N\gg1$) of mutually incoherent emitters. Then the net polarization information can be obtained by summing the Stokes parameters of each emitter element:
\begin{equation}
	I=\sum_i I_i,\quad	Q=\sum_i Q_i,\quad	U=\sum_i U_i,\quad	V=\sum_i V_i.
	\label{iquv}
\end{equation}
Many FRBs have been observed to exhibit extremely high degrees of polarization \citep{Michilli2018,Jiang2022,Zhang2023a,Xie2025,Wang2026}, which implies that different emitters may contribute nearly identical ellipticity angles and PAs; otherwise, depolarization would be expected. A similar treatment based on the incoherent superposition of radiation from multiple emitters or patches has also been discussed by \cite{Beniamini2025}.

\begin{figure*}
	\centering
	\includegraphics[width=0.8\textwidth]{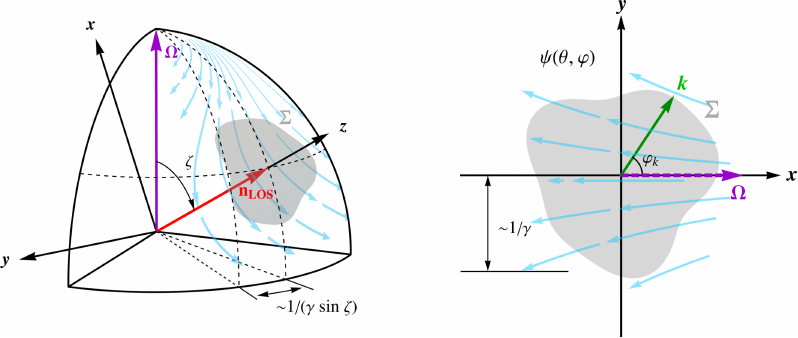}
	\caption{A three-dimensional Cartesian coordinate system defined on a unit sphere centered on the neutron star (left panel) and its two-dimensional projection onto the $x$-$y$ plane (right panel). The LOS unit vector $\boldsymbol{n}_\text{LOS}$ is along the $z$-axis, and the spin axis unit vector $\boldsymbol{\Omega}$ is in the $x$-$z$ plane. The angle between $\boldsymbol{\Omega}$ and $\boldsymbol{n}_\text{LOS}$ is $\zeta$. The gray area indicates the observable cone $\Sigma$ on the unit sphere. It has an angular size of order $1/\gamma$ and spans a phase width of order $1/(\gamma \sin\zeta)$, as seen from its projection onto the equatorial plane. The blue arrows denote the PA as a function of the angular coordinates, i.e., $\psi(\theta,\varphi)$. The vector $\boldsymbol{k}$ is defined as the gradient of $\psi(\theta,\varphi)$ at the point $(x,y,z)=(0,0,1)$.}
	\label{geneillus}
\end{figure*}

To facilitate a more detailed discussion, we introduce a Cartesian coordinate system on a unit sphere centered on a neutron star, as shown in Fig. \ref{geneillus}. An important assumption we require is that the PA of FRBs traces the geometric projection of some preferred direction in the emission region. That is, the PA can be described by a function $\psi(\theta,\phi)$, such that each pair of angular coordinates $(\theta,\phi)$ uniquely specifies a PA. In the framework of curvature radiation, for example, the direction of the PA is commonly related to the background magnetic field line at the emission site. More precisely, $\theta$ denotes the angle between the LOS and the plane of the curved magnetic field line, while $\varphi$ denotes the azimuthal angle of the plane relative to the LOS \citep{Rybicki1991}. Because of the beaming effect, radiation of an emitter is concentrated into a cone centered on the direction of its velocity, with a half-opening angle $\sim1/\gamma$. The radiation beam is observable only when the LOS lies within this cone, or equivalently, when the velocity direction of the emitter lies within a cone of half-opening angle $\sim 1/\gamma$ centered on the LOS, as illustrated in Fig. \ref{depoillus}. We therefore expect that equation (\ref{iquv}) can be written in the following continuous integrals:
\begin{equation}
	\begin{aligned}
		I&=\int_\Sigma w(\theta,\varphi)dS,\\
		Q&=\int_\Sigma w(\theta,\varphi)\cos{2\chi(\theta,\varphi,\lambda_1,\ldots,\lambda_n)}\cos{2\psi(\theta,\varphi)}dS,\\
		U&=\int_\Sigma w(\theta,\varphi)\cos{2\chi(\theta,\varphi,\lambda_1,\ldots,\lambda_n)}\sin{2\psi(\theta,\varphi)}dS,\\
		V&=\int_\Sigma w(\theta,\varphi)\sin{2\chi(\theta,\varphi,\lambda_1,\ldots,\lambda_n)}dS,
	\end{aligned}
	\label{constokeparam}
\end{equation}
where $w(\theta,\varphi)$ is a factor that weights the contributions of emitters with different directions to the total intensity. It is assumed that the ellipticity angle takes the general form $\chi(\theta,\varphi,\lambda_1,\ldots,\lambda_n)$, where $\lambda_1,\ldots,\lambda_n$ are additional parameters related to the radiation process. In general, the emission may be elliptically polarized, and unlike the PA, the ellipticity angle is not fully determined by the angular coordinates $(\theta,\varphi)$ alone \citep[e.g., ][]{Rybicki1991,Qu2023}. $dS=\sin{\theta}d\theta d\varphi$ is the infinitesimal surface element on the unit sphere. We define $\Sigma$ as the observable cone on the sphere: the emission from an emitter is observable only when its velocity direction lies within $\Sigma$, as shown in Fig. \ref{geneillus}. Its angular size should be of order $1/\gamma$. In general, $\Sigma$ is not necessarily a perfect spherical cap, since different emitters may have different Lorentz factors.

We retain the ellipticity angle in the formalism to keep the derivation applicable to a general polarization state, which is discussed first, before focusing mainly on the properties of linear polarization. Equation (\ref{constokeparam}) contains several uncertain parameters. We therefore begin with a discussion of two representative cases:
\begin{itemize}
	
	\item \textit{Case I: Slowly varying polarization field.} A well-studied case is when the polarization-related quantities $\chi(\theta,\varphi,\lambda_1,\ldots,\lambda_n)$ and $\psi(\theta,\varphi)$ vary slowly across the region $\Sigma$, so that they can be approximated by constant representative values within this region, i.e., $\chi\simeq\chi_0=\chi(\theta_0,\varphi_0,\lambda_1,\ldots,\lambda_n)$ and $\psi\simeq\psi_0=\psi(\theta_0,\varphi_0)$, for some $(\theta_0,\varphi_0)\in\Sigma$. This approximation is justified by the small angular extent of $\Sigma$. Equation (\ref{constokeparam}) then reduces to
	\begin{equation}
		\begin{aligned}
			I&\simeq\int_\Sigma w(\theta,\varphi)dS,\quad Q\simeq\cos{2\chi_0}\cos{2\psi_0}\int_\Sigma w(\theta,\varphi)dS,\\
			U&\simeq\cos{2\chi_0}\sin{2\psi_0}\int_\Sigma w(\theta,\varphi)dS,\quad V\simeq\sin{2\chi_0}\int_\Sigma w(\theta,\varphi)dS.
		\end{aligned}
	\end{equation}
	This indicates that the emission is highly polarized in this regime.

	\item \textit{Case II: Slowly varying radiation intensity.} Another important case is that $w(\theta,\varphi)$ varies slowly across $\Sigma$, so that the emission is approximately uniform within the region, at least on angular scales of order $1/\gamma$. For simplicity, we let $w(\theta,\varphi)\simeq w_0$. Applying the approximation to equation (\ref{constokeparam}), we immediately obtain 
	\begin{equation}
		\begin{aligned}
			I&\simeq w_0\int_\Sigma dS,\quad Q\simeq w_0\int_\Sigma \cos{2\chi(\theta,\varphi,\lambda_i)}\cos{2\psi(\theta,\varphi)} dS,\\
			U&\simeq w_0\int_\Sigma \cos{2\chi(\theta,\varphi,\lambda_i)}\sin{2\psi(\theta,\varphi)} dS,\\
			V&\simeq w_0\int_\Sigma \sin{2\chi(\theta,\varphi,\lambda_i)} dS,
		\end{aligned}\label{caseiistokes}
	\end{equation}
    where we have denoted $\chi(\theta,\varphi,\lambda_1,\ldots,\lambda_n)$ by $\chi(\theta,\varphi,\lambda_i)$ for convenience. Note that $\chi(\theta,\varphi,\lambda_i)$ and $\psi(\theta,\varphi)$ are not necessarily assumed to be constant within $\Sigma$ and may exhibit angular dependence. As a result, the polarization properties of the emission are not fully determined before the explicit integration over $\Sigma$. In the following, we therefore discuss the polarization properties in Case II in detail.
\end{itemize}

Equation (\ref{caseiistokes}) can be further simplified by assuming that all emitters share the same Lorentz factor $\gamma$, i.e., the region of integration $\Sigma$ can be approximately regarded as a spherical cap with a half-opening angle $1/\gamma$. We use this approximation and define a complex polarization parameter as 
\begin{equation}
	\mathcal{L}=Q+iU\simeq w_0\int_0^{1/\gamma}\theta d\theta\int_0^{2\pi}\cos{2\chi(\theta,\varphi,\lambda_i)} e^{2i\psi(\theta,\varphi)}d\varphi,
	\label{complexp}
\end{equation}
where $1/\gamma\ll1$ has been used. With this definition, the degree of linear polarization, circular polarization, and the PA are given by \citep{Rybicki1991}
\begin{equation}
	\Pi_\text{L}=\frac{|\mathcal{L}|}{I},\quad \Pi_\text{V}=\frac{V}{I},\quad\text{PA}=\frac{1}{2}\text{arg}(\mathcal{L}),
	\label{pilpa}
\end{equation}
where the function $\text{arg}$ denotes the argument of a complex quantity. We define
\begin{equation}
	V\simeq w_0 \int_0^{1/\gamma}V(\theta,\lambda_i)\theta d\theta,\quad \mathcal{L}\simeq w_0 \int_0^{1/\gamma}\mathcal{L}(\theta,\lambda_i)\theta d\theta,
\end{equation}
where 
\begin{equation}
	\begin{aligned}
		V(\theta,\lambda_i)&=\int_0^{2\pi}\sin{2\chi(\theta,\varphi,\lambda_i)}d\varphi,\\
		\mathcal{L}(\theta,\lambda_i)&=\int_0^{2\pi}\cos{2\chi(\theta,\varphi,\lambda_i)}e^{2i\psi(\theta,\varphi)}d\varphi.
	\end{aligned}
\end{equation}
By the Cauchy-Schwarz inequality:
\begin{equation}
	\left|\int_a^b g(\tau)d\tau\right|^2 \leq \left(b-a\right)\int_a^b \left|g(\tau)\right|^2 d\tau,
\end{equation}
we obtain
\begin{equation}
	V(\theta,\lambda_i)^2+\left|\mathcal{L}(\theta,\lambda_i)\right|^2\leq(2\pi)^2.
	\label{firstequality}
\end{equation}
The equality holds if and only if both $\chi$ and $\psi$ are independent of $\varphi$, i.e., $\chi(\theta,\varphi,\lambda_i)=\chi(\theta,\lambda_i)$ and $\psi(\theta,\varphi)=\psi(\theta)$. Next, we consider
\begin{equation}
	V^2+\left|\mathcal{L}\right|^2\leq\frac{w_0^2}{4\gamma^2}\int_0^{1/\gamma^2}\left[V(\theta,\lambda_i)^2 +\left|\mathcal{L}(\theta,\lambda_i)\right|^2\right]d\xi\leq I^2,
	\label{inequality}
\end{equation}
where $\xi=\theta^2$. The first inequality becomes an equality if and only if both $\chi$ and $\psi$ are independent of $\theta$, i.e., $\chi(\theta,\varphi,\lambda_i)=\chi(\varphi,\lambda_i)$ and $\psi(\theta,\varphi)=\psi(\varphi)$. We have used equation (\ref{firstequality}) in the second equality. Equation (\ref{inequality}) is straightforward, since the degree of polarization cannot exceed $100\%$. It follows that depolarization occurs when either $\chi$ or $\psi$ varies with $\theta$ or $\varphi$ within $\Sigma$.

The function $\chi(\theta,\varphi,\lambda_i)$ depends on the intrinsic radiation mechanism of FRBs, which remains poorly understood. In contrast, the function $\psi(\theta,\varphi)$ represents a preferred direction in the emission region, which, as assumed in this paper, may be determined by an extrinsic factor such as the background magnetic field. We thus suggest that there may be a degree of independence between these two functions. In this paper, we focus on the depolarization induced by a pure geometric effect. To avoid contamination from other parameters, we restrict our discussion to those FRBs with intrinsically low circular polarization. \footnote{It should be noted that some FRBs exhibit very strong circular polarization \citep[e.g., ][]{Jiang2024}. Therefore, the following discussion applies to bursts that are not intrinsically highly circularly polarized, although the observed high circular polarization may be caused by propagation effects.}  For simplicity, we let $\chi(\theta,\varphi,\lambda)\ll1$, so that equation (\ref{complexp}) reduces to
\begin{equation}
	\mathcal{L}\simeq w_0\int_0^{1/\gamma}\theta d\theta\int_0^{2\pi}e^{2i\psi(\theta,\varphi)}d\varphi.
	\label{simplifiedl}
\end{equation}
This is an oscillatory integral: if $\psi(\theta,\varphi)$ varies too rapidly, i.e., $\Delta\psi\gg1$ within $\Sigma$, then the degree of linear polarization $\Pi_\text{L}=\left|\mathcal{L}\right|/I\ll1$, which could lead to a nearly unmeasurable PA. Therefore, we focus on the case where $\psi(\theta,\varphi)$ exhibits significant but relatively slow variation, i.e., $\Delta\psi\sim1$ across the entire region $\Sigma$. We note that when $1/\gamma\ll1$, the coordinates transform as (see the right panel of Fig. \ref{geneillus}):
\begin{equation}
	x\simeq\theta\cos{\varphi},\quad y\simeq\theta\sin{\varphi}.
	\label{coordtrans}
\end{equation}
Within the small region $\Sigma$, we express $\psi(\theta,\varphi)$ as $\psi(x,y)$ and approximate it by its first-order expansion about the point $(x,y)=(0,0)$:
\begin{equation}
	\begin{aligned}
		\psi(\theta,\varphi)\longrightarrow\psi(x,y)&\simeq\psi_0+x\left(\partial_x\psi\right)_0 +y\left(\partial_y\psi\right)_0+\mathcal{O}(1)\\
		&=\psi_0+k\theta \cos{\left(\varphi-\varphi_k\right)}+\mathcal{O}(1).
	\end{aligned}
	\label{psifirstorder}
\end{equation}
The parameters $k$ and $\varphi_k$ are the length and azimuth angle of 
\begin{equation}
	\boldsymbol{k}=\left(\partial_x \psi, \partial_y \psi\right)_0,
\end{equation}
which is the gradient of $\psi(x,y)$ at the point $(0,0)$. By substituting the first-order expansion of $\psi(\theta,\varphi)$ into equation (\ref{simplifiedl}), we obtain
\begin{equation}
	\begin{aligned}
		\mathcal{L}&\simeq w_0 e^{2i\psi_0}\int_{0}^{1/\gamma}\theta d\theta \int_0^{2\pi}e^{2ik\theta \cos{\left(\varphi-\varphi_k\right)}}d\varphi\\
		&=w_0 e^{2i\psi_0}\frac{\pi}{k\gamma}J_1\left(\frac{2k}{\gamma}\right),
	\end{aligned}
	\label{simplifyp}
\end{equation}
where in the second line we have used the Bessel functions of the first kind:
\begin{equation}
	J_1(\eta)=\frac{1}{\eta}\int_0^\eta \tau J_0(\tau)d\tau,\quad J_0(\eta)=\frac{1}{2\pi}\int_0^{2\pi}e^{i\eta\cos{\tau}}d\tau.
\end{equation}
The quantity $2k/\gamma$ represents the total change in the PA across the region $\Sigma$. It can be related to the phase derivative of the PA through the relation:
\begin{equation}
	\frac{2k}{\gamma}\varepsilon\simeq\frac{d\text{PA}}{d\phi}\frac{2}{\gamma\sin{\zeta}},
\end{equation}
where $\sin{\zeta}$ arises from the projection of the opening angle of $\Sigma$ onto the equatorial plane of the neutron star, as illustrated in the left panel of Fig. \ref{geneillus}. The modulation factor $\varepsilon=\sin{\varphi_k}\lesssim 1$ is introduced from the projection of $\boldsymbol{k}$ onto the sweeping direction of the LOS (i.e., the $y$-axis shown in the right panel of Fig. \ref{geneillus}). In pulsars, it normally approaches unity when $d\text{PA}/d\phi$ reaches its peak value. The degree of linear polarization thus can be written as
\begin{equation}
	\Pi_\text{L}\simeq \left|\frac{2J_1(f)}{f}\right|,
	\label{pivsf}
\end{equation}
where the parameter $f$ acts as a proxy for the PA swing, and is given by
\begin{equation}
	f=\frac{d\text{PA}}{d\phi}\frac{2}{\varepsilon\gamma\sin{\zeta}}=\frac{d\text{PA}}{dt}\frac{P}{\pi\varepsilon\gamma\sin{\zeta}}.
	\label{definef}
\end{equation}
In the second equality, we have replaced $\phi=2\pi t/P$, with $P$ being the spin period of the neutron star. The validity of equation (\ref{pivsf}) is based on the assumption that $\zeta \gg 1/\gamma$, and it is no longer applicable for $\zeta \lesssim 1/\gamma$. Nevertheless, in this limit the LOS remains confined to the immediate vicinity of the spin axis, where it is geometrically difficult to produce a substantial PA swing while simultaneously maintaining a high degree of linear polarization. Therefore, we do not expect bursts in this regime to exhibit both properties at the same time.

\begin{figure}
	\centering
	\includegraphics[width=0.45\textwidth]{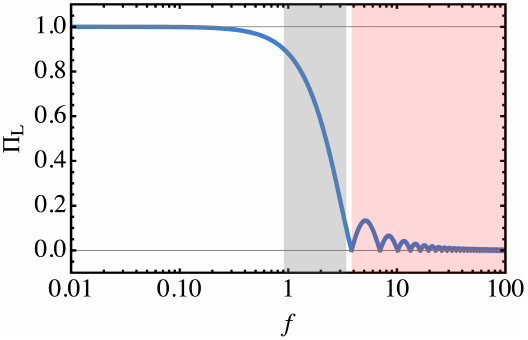}
	\caption{The degree of linear polarization $\Pi_\text{L}$, as a function of $f$, based on equation (\ref{pivsf}). Here, $f$ is a parameter characterizing the PA swing. The gray region indicates the range of $f$ where depolarization occurs and the PA remains measurable, corresponding to the transition from $\Pi_\text{L}=0.9$ to $\Pi_\text{L}=0.1$. The red region denotes the regime where the PA becomes essentially unmeasurable for $f \gtrsim 4$.}
	\label{pilfcurve}
\end{figure}

Equation (\ref{pivsf}) gives an approximate anti-correlation between $\Pi_\text{L}$ and $d\text{PA}/d\phi$ or $d\text{PA}/dt$. Fig. \ref{pilfcurve} displays the dependence of $\Pi_\text{L}$ on the parameter $f$. It shows that significant depolarization occurs when $f\sim1$, which is qualitatively consistent with our previous discussion in equation (\ref{criterion}). It also shows that the measurable PA swing is able to reach a maximum at $f\approx4$. Beyond this value, the degree of linear polarization drops sharply, and the PA is nearly unmeasurable. A tiny PA swing indicates $f\ll1$, and thus $\Pi_\text{L}\simeq1-f^2/8\to1$, which corresponds to the case of no geometric depolarization. We emphasize that the real relation between $\Pi_\text{L}$ and $d\text{PA}/d\phi$ depends on the detailed functional form of $\psi(\theta,\varphi)$. While our formulation is based on a general integral description of the polarization over the observable region, the analytic expression in equation (\ref{pivsf}) is obtained under a first-order expansion of $\psi$ around the LOS. In realistic systems, higher-order angular dependence may introduce additional corrections, making the exact relation more complex than the simplified form presented here. Physically, such complexity could arise from a non-dipolar magnetic field geometry (e.g., multipolar components), magnetospheric twisting, or physical variations across the emission region. Nevertheless, an anti-correlation between $\Pi_\text{L}$ and $d\text{PA}/d\phi$ may indeed exist. We therefore perform a preliminary test of this relation in pulsars in the next section.

\section{Test in Radio Pulsars}\label{testpulsar}

Radio pulsars provide an opportunity to test whether rapid PA swings can induce depolarization. We employ the pulsar sample compiled by \cite{Wang2023}, which provides a well-characterized dataset suitable for examining our conjecture. The sample includes polarization data $(\text{Bin},I, L, V, \text{PA})$ for a total of 682 pulsars, of which 190 were fitted using the RVM to derive their geometric parameters. \footnote{The pulsar polarization data used in this paper are available at \url{http://zmtt.bao.ac.cn/psr-fast/}.} To ensure high-precision PA measurements, we apply a uniform signal-to-noise threshold $(S/N>5)$ based on the linear polarization intensity, to filter reliable polarization data from the 190 pulsars. \footnote{Here, $S/N$ is defined as $L/\sigma_\text{L}$, where $L$ is the linear polarization intensity and $\sigma_\text{L}$ is its standard deviation estimated from the off-pulse data.} We then compute the degree of linear polarization and the derivative of the PA with respect to the pulsar phase for these pulsars. The degree of linear polarization is computed directly using $\Pi_\text{L}=L/I$ for each phase bin. The derivative of the PA at each phase bin $j$ is computed using a central difference scheme:
\begin{equation}
	\frac{d\text{PA}}{d\phi}\Big|_j\equiv\left|\frac{\text{PA}_{j+1}-\text{PA}_{j-1}}{\phi_{j+1}-\phi_{j-1}}\right|,
	\label{dpadphi}
\end{equation}
which provides a relatively stable profile for $d\text{PA}/d\phi$. The final $\Pi_\text{L}$ and $d\text{PA}/d\phi$ data are assembled into their corresponding phase profiles, which are then compared to examine any potential correlation between them.

\subsection{Identification of Anti-correlation}

In order to quantify the possible anti-correlation between the linear polarization fraction $\Pi_\text{L}$ and the PA swing rate $d\text{PA}/d\phi$, we compute the Spearman rank correlation coefficient $\rho_\text{s}$ between these two quantities. For each pulsar, we select only the PA data satisfying
\begin{equation}
	\frac{d\text{PA}}{d\phi}\geq0.3\text{ max}\left(\frac{d\text{PA}}{d\phi}\right).
\end{equation}
The choice of the coefficient $0.3$ is motivated as follows. Our theory suggests that depolarization becomes significant only when $d\text{PA}/d\phi$ is sufficiently large. Using the full dataset, which also includes small $d\text{PA}/d\phi$ values, would introduce a substantial number of irrelevant data points. From equation (\ref{pivsf}) and Fig. \ref{pilfcurve}, the onset of noticeable depolarization typically occurs at $\Pi_\text{L}(f_1)\approx0.9$, while the regime of strongest depolarization is reached at $\Pi_\text{L}(f_2)\approx0.1$. Under these approximate thresholds, we find $f_1/f_2\approx0.3$, which provides a natural scale for the selection criterion. This value therefore serves as an estimate of the range over which the anti-correlation is maintained.

\begin{figure*}
	\centering
	\includegraphics[width=1\textwidth]{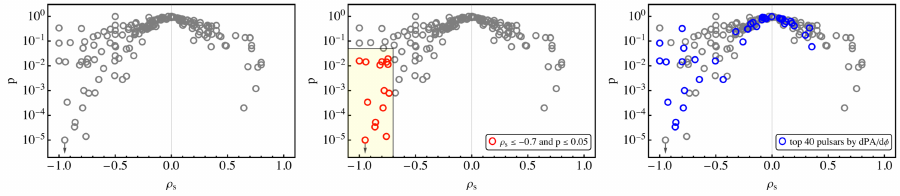}
	\caption{Scatter plot of Spearman rank correlation coefficients $\rho_\text{s}$ versus p-values for 181 pulsars with sufficient data ($N_ \text{Bin}>2$). Pulsars with $\rho_\text{s}\leq-0.7$ and $p\le0.05$ are marked in red (middle panel). The 40 pulsars with the largest $d\text{PA}/d\phi$ values are highlighted in blue (right panel).}
	\label{rhopdiagram}
\end{figure*}

\begin{figure*}
	\centering
	\includegraphics[width=1\textwidth]{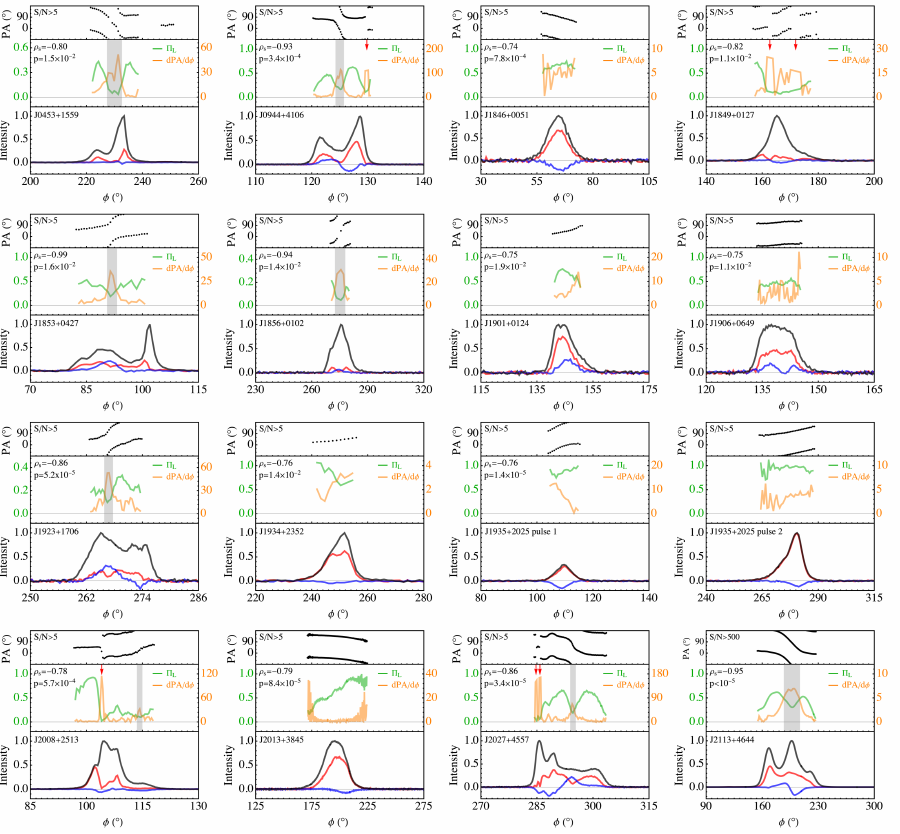}
	\caption{Polarization profiles of 15 pulsars identified as strongly anti-correlated candidates with $\rho_\text{s}\leq-0.7$ and $p\leq0.05$. In each subfigure, the lower panel shows the polarization profiles: the black, red, and blue curves represent the normalized total intensity, linear polarization intensity, and circular polarization intensity, respectively. The upper panel shows the PA profiles. The $S/N$ thresholds adopted for selecting the PA data are indicated in each subfigure. The middle panel compares the linear polarization fraction $\Pi_\text{L}$ with the PA swing rate $d\text{PA}/d\phi$, and the results of the correlation analysis between these two quantities are also shown. A subset of pulsars exhibits a visually apparent anti-correlation between these two quantities, which is highlighted using gray shaded regions. OPM jumps are marked with red arrows.}
	\label{pulsartest}
\end{figure*}

Among these 190 pulsars, 9 do not have sufficient data ($N_\text{Bin}\leq2$) for a correlation analysis. \footnote{The nine pulsars are J1839-0223, J1843+0119, J1855+0700, J1856+0912, J1903+0851g, J1913+0657, J1913+1050, J1928+1809g, and J1938+2659.} We therefore obtain the correlation coefficients $\rho_\text{s}$ and the corresponding p-values for the remaining 181 pulsars, as shown in Fig. \ref{rhopdiagram}. The Spearman rank correlation coefficient $\rho_\text{s}$ ranges from $-1$ to $1$, where $\rho_\text{s}=1$ indicates a perfect positive correlation, $\rho_\text{s}=-1$ indicates a perfect anti-correlation, and $\rho_\text{s}=0$ indicates no correlation between the two variables. The corresponding p-values quantify the statistical significance of the correlations under the null hypothesis of no intrinsic correlation between the two variables.

We empirically adopt $\rho_\text{s}\le-0.7$ and $p\le0.05$ as an initial criterion for statistically significant anti-correlation candidates. The impact of different $\rho_\text{s}$ thresholds will be further discussed in a later paragraph. Applying this criterion to the sample, we identify 15 pulsars that exhibit statistically significant anti-correlations between the linear polarization fraction $\Pi_\text{L}$ and the PA swing rate $d\text{PA}/d\phi$ (see the middle panel of Fig. \ref{rhopdiagram}). Their polarization profiles are presented in Fig. \ref{pulsartest}. Among these 15 candidates, 8 show a characteristic morphological behaviour, i.e., a local maximum in $d\text{PA}/d\phi$ that is approximately phase-aligned with a local minimum in $\Pi_\text{L}$. These visually apparent anti-correlations are highlighted by the gray shaded regions in Fig. \ref{pulsartest}. Some pulsars are also observed to exhibit orthogonal polarization mode (OPM) jumps. These events are characterized by abrupt $\sim90^\circ$ changes in PA, typically accompanied by local minima in the linear polarization fraction, and may therefore represent an extreme manifestation of depolarization associated with rapid PA variations. They are indicated by red arrows in Fig.~\ref{pulsartest}. The phase ranges associated with OPM jumps are not included in the computation of the correlation coefficients.

\begin{figure}
	\centering
	\includegraphics[width=0.4\textwidth]{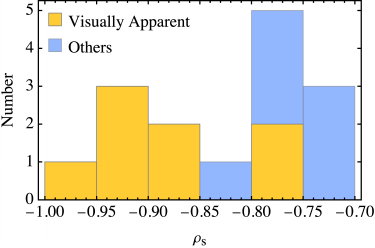}
	\caption{Distribution of the correlation coefficient $\rho_\text{s}$ for the 15 pulsars with $\rho_\text{s}\leq -0.7$ and $p\leq0.05$. The 8 pulsars with visually apparent anti-correlation have $-1 < \rho_\text{s} \lesssim -0.8$, while the remaining 7 pulsars have $-0.8 \lesssim \rho_\text{s} \leq -0.7$.}
	\label{rhobimdis}
\end{figure}

Notably, the $\rho_\text{s}$ values of the 8 pulsars with visually apparent anti-correlation are generally more extreme than those of the remaining 7 sources: the former have $-1 < \rho_\text{s} \lesssim -0.8$, whereas the latter have $-0.8 \lesssim \rho_\text{s} \leq -0.7$, as illustrated in Fig. \ref{rhobimdis}. One possible interpretation is that the 7 sources may represent spurious anti-correlations, as also suggested in Fig. \ref{rhopdiagram}, where a population of points around $\rho_\text{s} \sim 0.7$ indicates that similar clustering also appears on the positive side. This suggests that correlations with $|\rho_\text{s}| \lesssim 0.8$ may in part be driven by statistical fluctuations, and a more conservative criterion for identifying true anti-correlation may therefore be $\rho_\text{s} \lesssim -0.8$.

\begin{figure*}
	\centering
	\includegraphics[width=1\textwidth]{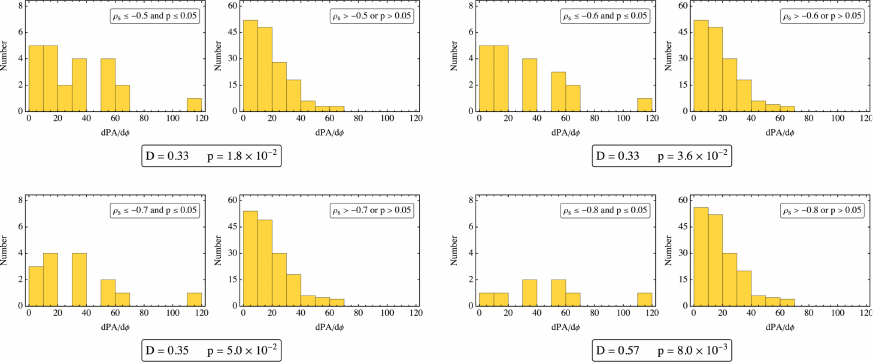}
	\caption{Distributions of \text{max}($d\text{PA}/d\phi$) for pulsars selected by different $\rho_\text{s}$ thresholds compared to the remaining population. KS test results are provided below each panel.}
	\label{padis}
\end{figure*}

We also highlight the 40 pulsars with the largest $d\text{PA}/d\phi$ values using blue markers in the right panel of Fig. \ref{rhopdiagram}. These 40 pulsars exhibit a pronounced asymmetric distribution in the diagram, which suggests a connection between large PA swings and the observed anti-correlation. We therefore apply different thresholds on $\rho_\text{s}$ to select sources with different levels of anti-correlation, and present the statistical distributions of $d\text{PA}/d\phi$ for these sources compared with the remaining sample, as shown in Fig.\ref{padis}. We further perform a two-sample Kolmogorov–Smirnov (KS) test to quantify the difference between the two populations, and the results are reported below each panel in Fig. \ref{padis}. The KS test statistic $D$ measures the maximum distance between the empirical cumulative distribution functions of the two samples, while the corresponding $p$-value quantifies the probability that the two samples are drawn from the same parent distribution. The results suggest that the distributions of $d\text{PA}/d\phi$ differ between the two populations, particularly for the case of $\rho_\text{s} \leq -0.8$. This provides further support for the connection between the observed anti-correlation and large PA swings.

\subsection{Lorentz Factor}

\begin{table}
	\centering
	\renewcommand{\arraystretch}{1.5}
	\caption{Key parameters of the 15 pulsars shown in Fig. \ref{pulsartest}. The last column indicates whether a clear visual anti-correlation between $\Pi_\text{L}$ and $d\text{PA}/d\phi$ exists.}
	\label{lorentzdata}
	\begin{tabular}{ccccc} 
		\hline
		Pulsar & $\text{max}(d\text{PA}/d\phi)$ & $\zeta\text{ }(^\circ)$ & $\gamma$ &  Visibility \\
		\hline
		J0453+1559 & 52.0 & $23.7^{+38.8}_{-14.4}$ & $258.8^{+384.9}_{-141.5}$ & Yes \\
		J0944+4106 & 114.2 & $133.2^{+40.05}_{-31.17}$ & $313.4^{+1630.6}_{-79.8}$ & Yes \\
		J1846+0051 & 8.2 & $152.8^{+32.2}_{-146.9}$ & -- & No \\
		J1849+0127 & 18.1 & $88.9^{+85.7}_{-87.8}$ & $36.3^{+1852.1}_{-0.007}$ & No \\
		J1853+0427 & 35.5 & $152.9^{+18.7}_{-52.5}$ & $155.7^{+329.9}_{-83.6}$ & Yes \\
		J1856+0102 & 31.0 & $24.6^{+73.3}_{-20.6}$ & $148.9^{+739.9}_{-86.9}$ & Yes \\
		J1901+0124 & 14.0 & $34.9^{+142.8}_{-41.8}$ & -- & No \\
		J1906+0649 & 11.2 & $142^{+35}_{-175}$ & -- & No \\
		J1923+1706 & 53.1 & $7.8^{+55.1}_{-5.9}$ & $782.2^{+2419.6}_{-662.9}$ & Yes \\
		J1934+2352 & 5.3 & $88.3^{+8.5}_{-7.2}$ & $10.7^{+ 0.13}_{-0.005}$ & No \\
		J1935+2025 & 12.6 & $80.8^{+1.7}_{-1.6}$ & $25.5^{+ 0.13}_{-0.11}$ & No \\
		J2008+2513 & 32.1 & $13^{+32.6}_{-9}$ & $285.8^{+635.8}_{-195.8}$ & Yes \\
		J2013+3845 & 35.0 & $72.1^{+35.3}_{-53.1}$ & $73.5^{+141.4}_{-3.6}$ & No \\
		J2027+4557 & 66.8 & $103.1^{+18.1}_{-20.1}$ & $137.1^{+19.0}_{-3.6}$ & Yes \\
		J2113+4644 & 6.9 & $179.553$ & $1779.2$ & Yes \\
		\hline
	\end{tabular}
\end{table}

\begin{figure*}
	\centering
	\includegraphics[width=1\textwidth]{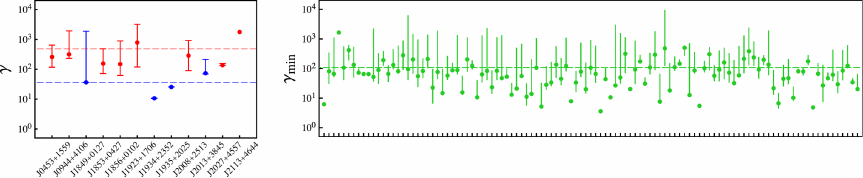}
	\caption{\textit{Left Panel}: Inferred Lorentz factors $\gamma$ for the 12 pulsars listed in Table \ref{lorentzdata}, derived from the anti-correlation between $\Pi_\text{L}$ and $d\text{PA}/d\phi$. Red points indicate sources with a visually identified anti-correlation, while blue points correspond to sources without such a clear anti-correlation. \textit{Right Panel}: Lower limits on $\gamma$ for the remaining sources with $\rho_\text{s} > -0.7$ or $p > 0.05$, under the assumption that the absence of anti-correlation is due to sufficiently large Lorentz factors. The dashed horizontal lines indicate the mean values of each color-coded population.}
	\label{gammadis}
\end{figure*}

The observed anti-correlation between $\Pi_\text{L}$ and $d\text{PA}/d\phi$ can be used to constrain the Lorentz factors of the particles responsible for pulsar radio emission. According to equation (\ref{definef}), the Lorentz factor is given by
\begin{equation}
	\gamma=\frac{d\text{PA}}{d\phi}\frac{2}{\varepsilon f \sin{\zeta}}.
	\label{gammaexp}
\end{equation}
The angle $\zeta$ and its uncertainty can be obtained from the geometrical relation $\zeta=\alpha+\beta$, where $\alpha$ is the magnetic inclination angle and $\beta$ is the impact parameter, both are inferred from RVM fits \citep{Wang2023}. \footnote{As the uncertainty in $\zeta$ is not provided, we estimate it by directly adding the uncertainties in $\alpha$ and $\beta$. This approach likely overestimates the true uncertainty, as it neglects the covariance between $\alpha$ and $\beta$.} By adopting a typical value $\varepsilon f \simeq 1$ and using the peak value of $d\text{PA}/d\phi$ as a representative estimate, we obtain approximate Lorentz factors for the pulsar sample. If the depolarization in the 15 pulsars shown in Fig. \ref{pulsartest} is indeed driven by rapid PA swings, the corresponding Lorentz factors can be estimated. The results are listed in Table \ref{lorentzdata} and shown in the left panel of Fig. \ref{gammadis}, where the 8 sources exhibiting visually apparent anti-correlation are distinguished for clarity. In three pulsars (J1846+0051, J1901+0124, and J1906+0649), the propagated uncertainty in $\zeta$ extends to values close to $0^\circ$ or $180^\circ$, for which $\sin{\zeta}$ approaches zero and equation (\ref{gammaexp}) becomes divergent. These sources are therefore not included from Fig. \ref{gammadis}. The derived Lorentz factors range from a few tens to over a thousand, broadly consistent with values reported in the literature \citep[e.g.,][]{Ruderman1975,Philippov2022}.

\begin{figure*}
	\centering
	\includegraphics[width=0.75\textwidth]{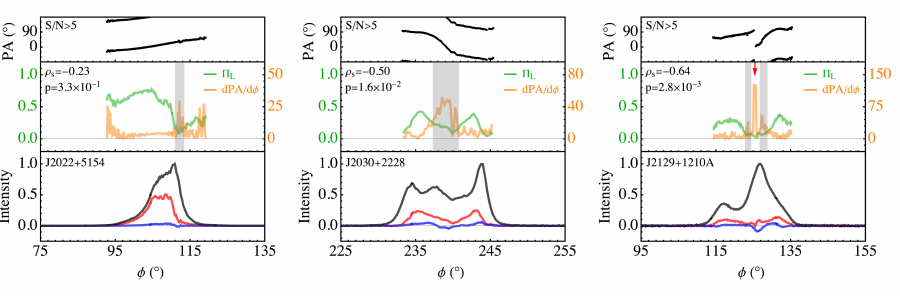}
	\caption{Polarization profiles of three representative pulsars, illustrating the impact of PA fluctuations on the correlation between $\Pi_\text{L}$ and $d\text{PA}/d\phi$. Although a visually apparent local anti-correlation can be seen, the measured correlation coefficients are not statistically significant, likely due to fluctuations in the PA data.}
	\label{unluckypulsartest}
\end{figure*}

Among these 181 pulsars, some exhibit relatively large $d\text{PA}/d\phi$ values but do not show a strong anti-correlation. One possible explanation is that these sources have relatively large Lorentz factors, which suppress depolarization. If this interpretation holds for all these sources without anti-correlation, then lower limits on their Lorentz factors can be derived, as shown in the right panel of Fig. \ref{gammadis}. For 57 of these sources, $\sin{\zeta}$ may approach zero within the propagated uncertainty, and they are omitted from Fig. \ref{gammadis}. An alternative possibility is that some of these sources may possess an intrinsic anti-correlation, but the measured correlation coefficients are weakened due to fluctuations in the PA data. We present three representative examples in Fig. \ref{unluckypulsartest} to illustrate this potential effect. In addition, other possibilities cannot be excluded. For instance, our theoretical model may be oversimplified for some sources, or the PA may be affected by additional propagation effects that are not fully accounted for in the current analysis \citep{Wang2015}.

\section{Applications in FRBs}\label{imfrbs}

If FRBs originate from rotating magnetospheres of neutron stars and are compatible with our paradigm, then according to equation (\ref{definef}), significant depolarization will occur when the time derivative of the PA reaches a critical value:
\begin{equation}
	\frac{d\text{PA}}{dt}\Big|_c\sim\left(20^\circ \text{ ms}^{-1}\right)\left(\frac{P}{1\text{ s}}\right)^{-1}\left(\frac{\varepsilon\gamma}{10^2}\right)\sin{\zeta},
	\label{criteriaint}
\end{equation}
where the spin period is set to $1\text{ s}$ to be consistent with typical Galactic magnetars \citep{Kaspi2017}. The Lorentz factor here has been assumed to be roughly comparable to that in a pulsar. In the following, we do not focus on a large, systematic sample of FRBs. Instead, we select a few well-studied sources from published papers as examples. These examples serve to demonstrate the preliminary application of our model in FRBs. A more comprehensive analysis of a larger population is left for future work.

\subsection{Possible Candidates?}

FRB 20221022A is of particular interest because of its nearly ideal S-shaped PA curve. We find that the degree of linear polarization seems to exhibit a subtle decline at the pulse center, where its PA changes most rapidly \citep[see][Fig. 1]{Mckinven2025}. If the depolarization is caused by the PA swing, then according to equation (\ref{criteriaint}), a period
\begin{equation}
	P\lesssim\left(0.2\text{ s}\right)\left(\frac{\varepsilon\gamma}{10^2}\right)\sin{\zeta}
\end{equation}
is inferred for its progenitor. Here, we have adopted an estimated value $d\text{PA}/dt\sim100^\circ \text{ ms}^{-1}$ from the observational data.

The second candidate, FRB 20210912A, is a one-off burst exhibiting an extremely steep PA swing over a very short time interval \citep{Bera2024,Bera2025}. It fails to maintain a high degree of linear polarization and even displays a slight dip in the linear polarization curve at the phase where its PA swing is most rapid (see \citealt[Fig. 3]{Bera2024}; \citealt[Fig. 5]{Bera2025}). If the PA swing is responsible for the dip, then one obtains
\begin{equation}
	P\lesssim\left(40\text{ ms}\right)\left(\frac{\varepsilon\gamma}{10^2}\right)\sin{\zeta},
\end{equation}
where we have estimated $d\text{PA}/d\phi\sim500^\circ \text{ ms}^{-1}$ from the observational data.

Interestingly, FRB 20181112A has a temporal profile very similar to that of FRB 20210912A \citep{Cho2020,Bera2024}. The difference is that it maintains a very high degree of linear polarization despite a large PA swing. We estimate a maximum of $d\text{PA}/dt\sim300^\circ \text{ ms}^{-1}$ for this burst \citep[see][Fig. 6]{Bera2024}. Substituting the estimated value into equation (\ref{criteriaint}) yields
\begin{equation}
	P<\left(60\text{ ms}\right)\left(\frac{\varepsilon\gamma}{10^2}\right)\sin{\zeta}.
\end{equation}
These results may imply that at least these one-off FRBs are produced by rapidly rotating neutron stars, rather than by magnetars with spin periods of a few seconds, like the progenitor of Galactic FRB 20200428 \citep{Bochenek2020,CHIME/FRBCollaboration2020,Lin2020,Zhu2023}.

\subsection{Spin-Magnetic Axis Alignment?}

Our theory allows us to constrain the angle between the LOS and the spin axis and to further examine whether active repeaters originate from neutron stars with aligned spin and magnetic axes \citep{Beniamini2025,Luo2025,Rajwade2025,Zhang2025}. 
Take the active repeater FRB 20240114A as an example, we adopt a maximum $d\text{PA}/dt\sim10^\circ\text{ ms}^{-1}$ \citep[see][Fig. 2, Burst 37]{Xie2025}. No significant depolarization indicates
\begin{equation}
	\sin{\zeta}> 0.5\left(\frac{P}{1\text{ s}}\right)\left(\frac{\varepsilon\gamma}{10^2}\right)^{-1}.
	\label{smazeta}
\end{equation}
Searches for the period of FRB 20240114A did not reveal any clear periodicity associated with neutron star spin \citep{Katz2025,Zhou2025}. Within the alignment scenario, the LOS always points toward the emission region, so that the periodicity is hidden. This may imply $\alpha+\zeta<\rho$, where $\alpha$ is the angle between the spin axis and the magnetic axis, and $\rho$ is the half-opening angle of the emission region. For an emission site $r_\text{em}$ far less than the light cylinder, the half-opening angle $\rho$ was estimated by \cite{Beniamini2025} as
\begin{equation}
	\rho\approx\left(0.07\text{ rad}\right)\left(\frac{P}{1\text{ s}}\right)^{-1/2}\left(\frac{r_\text{em}}{10^7\text{ cm}}\right)^{1/2}.
	\label{smarho}
\end{equation}
Combining equations (\ref{smazeta}) and (\ref{smarho}), we find this class of models is restricted to a rather fine-tuned region of parameter space. Searching for similar signals in the future may provide tighter constraints on the magnetic inclination angle.

\section{Discussion}\label{discussion}

We have proposed that rapid PA swings may induce depolarization in FRBs, provided that the PA has a geometric origin. We tested this conjecture in a pulsar sample and may have observed the expected anti-correlation between the degree of linear polarization and the derivative of the PA with respect to spin phase in a subset of sources. This allows us to estimate the Lorentz factors of the particles responsible for their radio emission. We also attempted to apply this theory to constrain the spin periods of FRB progenitors. Our tentative results suggest that the progenitors of at least some one-off FRBs may be fast-spinning neutron stars with sub-second spin periods. Furthermore, we highlight the potential of our theory to examine the scenario in which repeaters are powered by neutron stars with nearly aligned spin and magnetic axes \citep{Beniamini2025,Luo2025,Rajwade2025,Zhang2025}. Taking the active repeater FRB 20240114A as an example, our analysis suggests that it is somewhat difficult for the neutron star responsible for this source to have a small magnetic inclination angle. Therefore, the absence of periodicity associated with neutron star rotation in FRBs may require alternative explanations, such as multiple stochastic emission sites \citep{Zhu2023,Du2025,Manaswini2026}. However at present, it is still too early to draw any definitive conclusion. It would be worthwhile for future work to systematically search for the anti-correlation between the linear polarization fraction $\Pi_\text{L}$ and the PA swing rate $d\text{PA}/dt$ in large FRB samples.

One of the major uncertainties in the application of this theory to FRBs comes from the Lorentz factor of the emitting particles. We have to assume $\gamma\sim100$, a value similar to that inferred from pulsars. While the Lorentz factor can in principle be calculated from some specific models in FRBs \citep{Dai2016,Wang2016}, a method to estimate it should be developed in the future. Another uncertain parameter is the spin period. It is possible that the rapid perturbations (with characteristic timescales of $\sim\text{ms}$) of the local magnetic field lines during FRB emission phases are misinterpreted as global magnetospheric rotation. In this case, the adopted period may not reflect the intrinsic spin period of the neutron star. We emphasize that our theory is capable of testing this scenario if we are fortunate to have enough instances of depolarization caused by PA swings. According to equation (\ref{criteriaint}), if the misinterpreted period varies over time (overwhelms the variation of the Lorentz factor), temporally evolving $d\text{PA}/d\phi\big|_c$ should be observed.

Depolarization caused by PA swings may occur in FRBs when the instrument has a finite temporal resolution \citep{Beniamini2025}. However, this effect is generally too weak to compare with the one discussed in this paper. A relativistic magnetized shock model may also predict a relation between the degree of linear polarization and PA swings \citep{Iwamoto2024}. In contrast, our model requires that significant depolarization occurs only when the PA varies rapidly, which appears to be more consistent with observations. In pulsars, rapid PA swings have also been suggested to be associated with emission from lower altitudes, where the radiation may have a lower linear polarization fraction or may be affected by the superposition of OPMs \citep{Wang2015}. Such effects could potentially contribute to the observed anti-correlation between $\Pi_\text{L}$ and $d\text{PA}/d\phi$ and should therefore be carefully disentangled from geometric depolarization. Future work should investigate the frequency dependence of the geometric depolarization and test the anti-correlation on a single-pulse basis.

\section{Acknowledgments}

We thank an anonymous referee for valuable comments and constructive suggestions that have allowed us to improve our manuscript. Y.-C. Huang thanks Yi-Xuan Shao, Fa-Yin Wang, Ren-Xin Xu, Wei-Yang Wang, Yi-Nan Chen, Jia-Pei Feng, Mi-Xiang Lan, and Biao Zhang for useful discussions. The authors acknowledge the use of ChatGPT (OpenAI), DeepSeek, and Gemini (Google) for assistance with English language editing. This work was supported by the National Natural Science Foundation of China (grant No. 12393812) and the Strategic Priority Research Program of the Chinese Academy of Sciences (grant No. XDB0550300).

\section{Data Availability}

The polarization data of pulsars are available at \url{http://zmtt.bao.ac.cn/psr-fast/}.

\bibliography{MPE}
\bibliographystyle{mnras}

\bsp	
\label{lastpage}
\end{document}